# A Hybrid Model for Estimating Software Project Effort from Use Case Points

Mohammad Azzeh
Department of Software Engineering
Applied Science University
Amman, Jordan POBOX 166
m.y.azzeh@asu.edu.jo

Ali Bou Nassif
Department of Electrical and Computer Engineering
University of Sharjah
Sharjah, UAE
anassif@sharjah.ac.ae

**Abstract.**
Early software effort estimation is a hallmark of successful software project management. Building a reliable effort estimation model usually requires historical data. Unfortunately, since the information available at early stages of software development is scarce, it is recommended to use software size metrics as key cost factor of effort estimation. Use Case Points (UCP) is a prominent size measure designed mainly for object-oriented projects. Nevertheless, there are no established models that can translate UCP into its corresponding effort; therefore, most models use productivity as a second cost driver. The productivity in those models is usually guessed by experts and does not depend on historical data, which makes it subject to uncertainty. Thus, these models were not well examined using a large number of historical data. In this paper, we designed a hybrid model that consists of classification and prediction stages using a support vector machine and radial basis neural networks. The proposed model was constructed over a large number of observations collected from industrial and student projects. The proposed model was compared against previous UCP prediction models. The validation and empirical results demonstrated that the proposed model significantly surpasses these models on all datasets. The main conclusion is that the environmental factors of UCP can be used to classify and estimate productivity.

## 1. Introduction

Use Case Points (UCP) is a well-established software sizing technique that utilizes a UML use case diagram to estimate the size of object-oriented projects at the early stages of software development [2][16]. The basic idea of UCP was mainly inspired from another software sizing technique that depends mainly on functional requirements, called Function Points [1][24]. Karner [16] established a well-defined procedure to convert the use case diagram elements into a set of metrics that reflect the work effort needed to accomplish software projects. The translation procedure requires a standard of writing use case descriptions. Thus, it is recommended to avoid the free style that often depends on natural language and follow a common guideline [24]. The first version of UCP lacks validation and examination about its reliability for software organizations. The major challenge in UCP is the arbitrary numbers involved in calculating the software size. In fact, there is no justification as to how these numbers were found. In addition, there is no efficient method that can convert the UCP size into its corresponding software effort in terms of person-hours or person-months. Therefore, it was hard to build an effort prediction model because of the limitation in the number of collected projects. The first version of effort estimation based on UCP suggests the use of productivity as a second cost driver, as shown in Equations 1 and 2. This approach has long been used in many studies conducted on UCP, but the validity of this approach has not been well examined over a large number of observations.

Software productivity is defined as a ratio between effort and size [12][14]. This relationship has two contradicting interpretations. On one hand it can be defined as project productivity when it is measured as effort/size. On the other hand it can be defined as team productivity when it is measured as size/effort. Both definitions are used within the software engineering community, but the first one is preferable. The effort estimation model is usually constructed based on the productivity interpretation. Equation 1 is used to compute effort when the productivity ratio is interpreted as project productivity. In contrast, Equation 2 is used when the productivity ratio is interpreted as team productivity [2]. Nevertheless, computing the software productivity must be made before the effort can be estimated. This may depend on many variables such as: reuse percentage, type of software process, team communications, and the number of deliverables. Although adopting good development practices may increase the productivity, it does not always do so because of circumstances outside the control of the software development team.

$$Effort = Productivity \times UCP \qquad (1)$$

$$Effort = \frac{\alpha}{Productivity} \times UCP^{\beta} \qquad (2)$$

Estimating effort from UCP can fall into one of three approaches. The first approach presumes that there are no historical projects available within the software organization so the project manager must pre-determine the productivity ratio for the software project. In this case, the decision is heavily dependent on the estimator and subject to a large degree of bias. Typical examples that have followed this approach are the studies conducted by Karner [16] and Schneider and Winters [18]. Karner [16] proposed a fixed productivity ratio (=20 person-hours/UCP) for all software projects. This approach is not practical because it does not take into consideration the type, complexity, domain, and environments of a software project. In contrast, Schneider and Winters [18] proposed three levels of productivity (fair=20, low=28, and very low=36) person-hours/UCP based upon analysing the environmental factors of a software project. The second approach uses machine learning and data mining techniques to build regression models that exhibit the relationship between effort and UCP. This approach does not need to pre-determine the productivity but needs historical data to build the regression model. Nevertheless, this approach is affected by the number of projects in the training set, setup parameters, and validation procedure. The third approach attempts to use both of the above approaches in one model. An example of this scenario is the work proposed by Nassif et al. [2], who proposed four levels of the productivity ratio based on the weighted sum of the environmental factor and using an expert-based fuzzy model. Nassif et al. [2] built a log-linear regression model that uses UCP and productivity.

Above all, we can see that all previous models were constructed using a very limited number of observations. In addition, the assumptions made about using productivity ratios have not been well examined. Using fixed or limited productivity ratios did not contribute well to improving prediction accuracy. No previous studies attempted to study the relationship between productivity and environmental factors when historical data was available. In fact, the productivity prediction should be flexible and adjustable when historical data is present. The flexibility means that the productivity must be affected by the UCP factors assessment. The adjustability means the productivity of one project should be adjusted based on the productivity from the historical projects. Finally, there is no study that has attempted to examine the effect of using UCP components with productivity to predict effort.

Stimulated by this situation, we proposed a new effort estimation model that can support management decisions during the feasibility study and project inception. The proposed model consists of two stages. In the first stage, the historical productivity is clustered to create fine-grained productivity labels and then classified based on environmental factors. For that purpose we used the bisecting k-medoids clustering technique [19][22] and support vector machine [11]. The predicted productivity, computed during the test phase, is based on the centre of the predicted productivity label. The studies conducted by [2] and [18] showed that the environmental factors can work as good indicators for software productivity since they reflect the team workload within the software project. In the second stage, the effort estimation model is built using Radial Basis Neural Network (RBFNN) [15]. The model is trained using historical UCP and productivity variables. Then during the estimation process, the predicted productivity from stage one is entered with the UCP of the new project as input to RBFNN to predict effort. The proposed model has advantages over previous models in that it can learn productivity from environmental factors using classification and decomposition techniques. So the number of productivity levels in each training set depends on the structure of that set. It also offers a non-linear learning mechanism to mimic the relationship between effort and two other predictors (UCP and productivity)

The rest of this paper is organized as follows: Section 2 gives an introduction to Use Case Points. Section 3 presents related work. Section 4 introduces the proposed model. Section 5 presents research methodology. Section 6 shows the empirical results and discussion. Section 7 presents threats to validity and, finally, Section 8 presents conclusions.

## 2. An overview of Use Case Points

The UCP estimation method was first introduced by Karner [16] to predict the size of object-oriented software projects. The UCP is computed by converting the elements of UML use case diagram to size metrics through a well-defined procedure. In the first step, the estimator must classify the actors in the use case diagram into three categories according to their difficulties: simple, average, and complex, as shown in Table 1. Based on that the Unadjusted Actor Weights (UAW) is computed, as shown in Equation 3. Similarly, the use cases are also classified into three classes (simple, average, and complex) based on the number of transactions mentioned in the use case descriptions, shown in Table 2. A transaction is defined as a stimulus and response occurrence between the actor and the system [21]. Based on that, the UUC is calculated as shown in Equation 4. The Unadjusted Use Case Points (UUCP) is computed based on the summation of UAW and UUC.

Table 1. Types of Actors

| Type | Description |
|---|---|
| Simple | Actor interacts using API |
| Average | Actor interacts using text-based interface |
| Complex | Actor interacts using Graphical User Interface |

$$UAW = 1 \times sa + 2 \times aa + 3 \times ca \qquad (3)$$

Where $sa, aa, ca$ are the numbers of simple, average, and complex actors respectively.

Table 2. Types of Use Cases

| Type | #transactions |
|---|---|
| Simple | <=3 |
| Average | 4 to 7 |
| Complex | >7 |

$$UUC = 5 \times suc + 10 \times auc + 15 \times cuc \qquad (4)$$

Where $suc, auc, cuc$ are the numbers of simple, average, and complex use cases respectively.

$$UUCP = UAW + UUC \qquad (5)$$

Finally, the UUCP should be adjusted by two sets of adjustment factors: Technical Complexity Adjustment Factor (TCF) and Environmental Adjustment Factor (EF). TCF is computed from a set of 13 technical factors ($F_1$, $F_2$, ..., $F_{13}$) that have great influence on project performance. Similarity, EF is computed from a set of eight environmental factors ($E_1$, $E_2$, ..., $E_8$) that have great effect on productivity. Each factor in both sets can take an influence value between zero and five and predefined weights that reflect the influence of that factor. Equations 6 and 7 demonstrate how TCF and EF are calculated, respectively. Finally, the UCP is computed based on Equation 8. More details about these factors can be found in [6].

$$TCF = 0.6 + \left(0.01 \times \sum_{i=1}^{13} (F_i \times fw_i)\right) \quad (6)$$

where $F_i$ is the value of influence of factor $i$, and $fw_i$ is the weight associated with factor $i$.

$$EF = 1.4 - \left(0.03 \times \sum_{i=1}^{8} (E_i \times ew_i)\right) \quad (7)$$

where $E_i$ is the value of influence of factor $i$, and $ew_i$ is the weight associated with factor $i$.

$$UCP = UUCP \times TCF \times EF \quad (8)$$

3. Related Work

All studies on UCP can be divided into three directions, the studies in the first direction attempted to improve the construction of the UCP sizing technique [33]. In this context, Braz et al. [26] proposed an improvement on the classical UCP by using fuzzy numbers. The new metric is called Fuzzy Use Case Size Points that enables gradual classifications in the estimation. Robiolo et al. [17] introduced two variables, which express a notion of the size-transactions and entity objects. These two variables can be calculated from a description of the use case. On the other hand, Mohagheghi et al. [29] made a significant change on the original UCP method in order to adapt it for incremental development by modifying the complexity assessment of actors and handling the non-functional requirements. The authors figured out that the accuracy of UCP depends on many factors set empirically, and they suggest that the value of person-hours needs to be increased for larger-sized projects. Azzeh [23] added a new adjustment factor to manage UCP with global software project development.

The studies in the second direction focus on how to build effort estimation based on UCP by using machine learning and data mining techniques [3][4][20]. In this regard, Nassif et al. [2] presented a log-linear regression model which consists of building a relationship among effort, UCP, and productivity. A fuzzy logic approach based on the value of Environmental Adjustment Factors was used to find out the productivity ratio. Also, a multilayer perceptron neural network model was built to estimate effort from UCP and team productivity. A Treeboost model [7] was used to estimate effort based on UCP and team productivity. Neural network based models [5][6][31] have also been used to predict the software effort. The inputs of these neural network models include software size in UCP and other quality attributes such as complexity. Pantoni et al. [30] used fuzzy set theory to predict the software effort, which could be a good alternative to other techniques given the absence of historical data. However, in the long term estimation, other techniques give more accurate results.

The studies in the last direction aim to examine and simplify UCP in order to enhance the precision of estimation. In this regards, Anda et al. [9][10]investigated the accuracy of the UCP model and found that the estimates rely heavily on the technical and environmental factors. The authors pointed out that the UCP can be very helpful for those who wish to predict effort from UML use cases, but are not expert in their field. In addition, the authors recommend that the UCP model can be tailored by calibrating the environmental factors based on the type of the organization. Ochodek et al. [24][25] attempted to simplify the process of calculating the UCP by excluding some UCP components because they are insignificant with respect to the effort estimation.

Indeed, we can observe that none of the above studies used a hybrid model to predict productivity and effort from historical data at the same time. The proposed model can learn from historical data to obtain finer grained productivity labels, and then predict productivity and effort.

## 4. The proposed Model

In this section we discuss the implementation of the proposed effort estimation model. This paper mainly focuses on two important issues related to effort estimation based upon UCP. The first issue is related to predicting productivity from eight environmental factors ($E_1, E_2, ..., E_8$), and the second issue is to estimate effort from the predicted productivity and UCP. Therefore, the proposed prediction model is a hybrid model that consists of two stages to cope with the abovementioned issues. As mentioned in Section 2, there is a strong relationship between environmental factors and project productivity because each factor has a significant impact on the project productivity. For example, the first factor, which is "Familiar with Rational Unified Process (RUP)," evaluates the familiarity with RUP, so higher experience with RUP leads to better productivity and less effort. Another example is the "Lead Analyst capability" factor that reflects the capability and knowledge of the developer, so better capability means better productivity. Therefore, we believe that all environmental factors could be good indicators of project productivity. Next we will discuss the construction of the proposed model during training and testing stages.

### 4.1 Training Phase

During the training phase we construct two prediction models: one for productivity and one for effort. In the next subsections we will discuss the implementation of both models.

### 4.1.1 Productivity Prediction

The eight environmental factors ($E_1, E_2, ..., E_8$) are used to classify and predict productivity. Since each environmental factor exhibits the ordinal scale (i.e., each factor can take a distinct value between 0 and 5), we decided to build a classification model. But first we need to convert productivity variable from a numerical scale to the nominal scale, which represents levels of productivity. To accomplish that we applied a clustering technique on the productivity to create coherent and finer labels. The unsupervised learning method called k-medoids [19] has been used for that goal. The k-medoids is a centroid-based clustering method that divides data into $C$ distinct groups. The major distinction between k-medoids and conventional k-means is the choice of cluster centers. In k-means method the center is calculated based on the average of instances within a class, but in k-medoid the center is identified based on the most centrally located instance in the cluster [19]. The popularity of making use of k-medoids clustering comes from its ability to use arbitrary dissimilarity or distance functions, which also makes it an appealing choice of a clustering method for software effort data as software effort datasets often exhibit very dissimilar characteristics.

    Like k-means, the k-medoids require determining the number of optimal clusters before performing a clustering procedure. This step is usually performed based on guessing or using clustering validity indexes. Both techniques do not frequently produce the optimal number of clusters. Therefore, we used the bisecting procedure with k-medoids to avoid arbitrarily assigning the number of clusters. The bisection k-medoids [19][22] work by recursively applying the basic k-medoids algorithm and splitting each cluster into two sub-clusters to form a binary tree of clusters, starting from the whole dataset. This enables us to better understand

the structure of the dataset and helps to create coherent and finer classes. The algorithm starts with considering the whole dataset as the initial cluster. Then for each step, one cluster is bisected into two coherent clusters, as shown in Figure 1. The stoppage criteria depends on the comparison between the variance of the childes' cluster and their direct parent in the tree. The variance within cluster is calculated as shown in Equation 9; a smaller value shows a high homogeneity (i.e., less scattering). If the variance of the parent cluster is much smaller than the largest variance of both child clusters then the clustering procedure is stopped, otherwise the algorithm continues clustering. The complete bisecting k-medoid algorithm is shown in Figure 2.

$$variance = \frac{1}{n}\sum_{j=1, y_j \in C_i}^{n} \|x_j - v_i\|^2 \qquad (9)$$

where $\|\cdot\|$ is the usual Euclidean norm, $x_j$ is the $j^{th}$ data object and $v_i$ is the center of $i^{th}$ cluster ($C_i$).

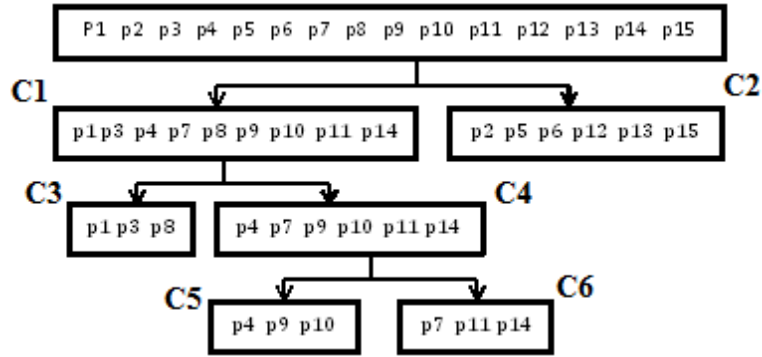

Figure 1 illustration of Bisecting k-medoids algorithm [22]

```
1   Input: The dataset X
2   Output: The set of N clusters S={C₁, C₂, C₃, C₄, ..C_N}
3   Initialization: Let V=X, S={}, NextLevl={}
4   Repeat while size(V)> 0
5     foreach Cluster C in V
6       Comp←variance (C)
7       [C₁,C₂]← k-medoids(C,2)
8       Comp1←variance(C₁)
9       Comp2←variance(C₂)
10      If(max(Comp1,Comp2)<Comp)
11        NextLevel←NextLevel∪{C₁,C₂}
12      Else
13        S←S∪{C}
14    End
15    V←NextLevel
16    NextLevel←{}
17  End
```

Figure 2 Bisecting k-medoids algorithm [22]

The leafs obtained are considered the productivity labels, so each label is given an appropriate linguistic name such as: fair, low, very low…etc. Then the training dataset (including eight environmental factors as input and discovered productivity labels as output) are entered into the support vector machine algorithm (SVM) to learn productivity from environmental factors. The SVM is used because it has been proven to be one of the most accurate classification algorithms in literature [11]. SVM is a binary classifier defined by a separating hyperplane. The fundamental notion of SVM is that given training data $\{(x_1, y_1), \ldots, (x_n, y_n)\}$ where $x \in \Re^m$ is the space of input pattern, and $y \in \Re$ is the output variable, the SVM generates a hyperplane that separates the data points into two classes with maximal margin. Margin is defined as the maximal width of the slab parallel to the hyperplane that has no interior data points. Among different Kernel functions that are used within SVM, we used the nonlinear Gaussian function. Also we used the sequential minimal optimization method to find the best hyperplane. The memory consumption is controlled by the value that specifies the kernel matrix cache with 5000. The Iteration limit was set to 1000, epsilon value was set to 0.001. The LIBSVM [35] tool that is implemented in MATLAB has been used to build our SVM model.

### 4.1.2 Effort Prediction

The effort estimation model is constructed using RBFNN. The inputs for this model are UCP size measure and productivity of training projects, as shown in Figure 3. It is important to note here that the productivity variable is the actual numerical values, not the labels that are used in the previous step. The RBFNN is a feed-forward network which consists of three layers: input, hidden, and output [15][34]. The neurons in the hidden layer include a non-linear activation function (mainly a Gaussian function), whereas the output layer includes linear function.

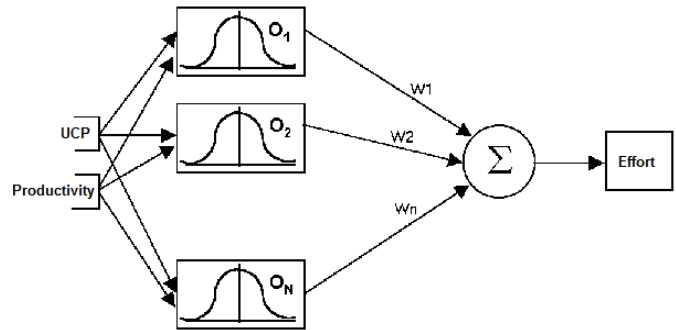

Figure 3: Radial Basis Function Neural Network

The Gaussian function in each neuron uses the distance between the input data and its center. Each neuron may have different Gaussian function with a different spread and center, which calculates the distance between the training input vector and its prototype vector. The value of the kernel function is closer to one when the input vector is very similar to the neuron prototype. The RBFNN model is trained based on the algorithm proposed by Chen et al. [32], which is an orthogonal forward selection algorithm based on the leave one-out criterion. The number of the neurons in the hidden layer is variable and is affected by the structure of training data. In this paper, the hidden neurons are added to the network one by one until the optimal training results are achieved (i.e., minimum means squared errors). The best results are obtained when the errors in the training stage and the validation stage are minimal. The hidden neurons in Dataset1, Dataset2, and Dataset3 are 5, 6, and 8, respectively. The default spread value (i.e., $\sigma_i = 1$) is used for all added hidden neurons to avoid any possible effect for a small or large spread.

$$f(x) = e^{\left(\frac{(X-\mu_i)^2}{2\sigma_i^2}\right)} \tag{10}$$

Where $\mu_i$ is the center and $\sigma_i$ is the spread of the $i^{th}$ hidden neuron.

### 4.2 Testing Phase

During the testing phase, all test observations must have the same structure of training observations. First, we enter the eight environmental factors of the test case into the SVM classification model to predict the productivity label. The medoid value of the corresponding label (i.e., cluster) will be used as the productivity ratio for that test project. Next, we enter the predicted productivity from the previous step along with its UCP into the constructed RBFNN model. The outcome of the RBFNN is the estimated effort for the test project.

## 5. Methodology

In this section we discuss various aspects of our experimental study as follows: validation, datasets, and evaluation measures.

### 5.1 Validation and Experiment Methodology

To validate all prediction models we used leave-one-out cross validation (LOOCV). The procedure of LOOCV works in this way: in every step, one project is taken as a test set and the rest of the projects act as training sets. The training dataset is used to construct both the productivity classification algorithm and the RBFNN effort estimation model. The test case is entered into the classification algorithm, then the predicted productivity along with UCP components are entered into the RBFNN model to produce the effort estimate. The accuracy measures are calculated for every test set. The distinct advantage of using LOOCV over other n-fold cross validations is that it is used in a deterministic procedure that can be exactly repeated by any other research with access to a particular dataset. Furthermore, it ensures that any prediction model is constructed from the same set of training projects. Based on literature, the LOOCV produces lower bias estimates and higher variance than the n-Fold cross validation, because the method needs to learn from a large number of examples and conduct more tests.

To ensure the performance of the proposed model, we made comparisons with previous well-known effort estimation models that use both UCP and productivity as effort drivers, such as Karner [16], Schneider and Winters (S&W) [18], and Nassif et al. [2]. Karner [16] is the first version of UCP effort estimation which uses a fixed productivity ratio (20 person-hours) for all projects regardless of their types, complexity, and environments. Equation 11 illustrates the model of Karner.

$$Effort = 20 \times UCP \tag{11}$$

S&W [16] uses the same model as Karner but with three productivity ratios: fair (20 person-hours/UCP), low (28 person-hours/UCP), and very low (36 person-hours/UCP). Finding the appropriate ratio depends on analyzing environmental factors. Basically, one should count the number of factors that have influence values of less than three from the set ($E_1$ to $E_6$) and count the number of factors that have influence value larger than three from the set ($E_7$ to $E_8$). Based on the total count, the productivity is evaluated as shown in Equation 12.

$$Effort = \begin{cases} 20 \times UCP & total\_Count \leq 2 \\ 28 \times UCP & 3 \leq total\_Count \leq 4 \\ 36 \times UCP & total\_Count > 4 \end{cases} \tag{12}$$

Nassif et al. [2] developed a log-linear regression model based upon UCP and productivity, as shown in Equation 13. In the original model, the values were α=8.16 and β=1.17. In this paper, we follow the same procedure conducted by Nassif et al. [2] but the value of α and β will be computed based on the training dataset in each validation run. The productivity is computed based on finding $prod\_sum = \sum_{i=1}^{8}(E_i \times ew_i)$ from the environment factors. Then this value is entered into their proposed fuzzy model that converts the *prod_sum* into the productivity value using some predefined rules. The productivity ratio is defined into four fuzzy sets representing four levels of productivity ratios.

$$Effort = \frac{\alpha}{Productivity} \times UCP^{\beta} \qquad (13)$$

### 5.2 Datasets

In this paper we used two datasets that contain a reasonable number of projects in comparison with previous studies. The first dataset (Dataset1) contains 45 industrial projects collected from software organization. The architecture of these projects was two-tier and three-tier applications [2]. The second dataset (Dataset2) contains 65 educational projects that were collected from 4th year and Master's university students. The projects have been designed and coded using UML diagrams and OO programming languages [2]. Since both datasets have the same structure, we merged Dataset1 and Dataset2 into one dataset, called third dataset (Dataset3). This enables us to investigate the usefulness of the proposed model over the heterogeneous dataset. Tables 3, 4, and 5 demonstrate the summary statistics for three datasets.

TABLE 3. Descriptive statistics of Dataset1

| Variable | Mean | StDev | Skewness | Kurtosis |
|---|---|---|---|---|
| UCP | 739.3 | 1563.9 | 3.0 | 11.7 |
| Effort | 20573.5 | 47326.9 | 3.2 | 12.4 |
| Productivity | 24.1 | 5.1 | 0.0 | 2.2 |

TABLE 4. Descriptive statistics of Dataset2

| Variable | Mean | StDev | Skewness | Kurtosis |
|---|---|---|---|---|
| UCP | 82.6 | 20.7 | 0.8 | 4.1 |
| Effort | 1672.4 | 414.3 | -0.1 | 2.2 |
| productivity | 20.8 | 4.8 | 0.2 | 2.7 |

TABLE 5. Descriptive statistics of Dataset3

| Variable | Mean | StDev | Skewness | Kurtosis |
|---|---|---|---|---|
| UCP | 351.3 | 1045.3 | 5.1 | 30.1 |
| Effort | 9404.7 | 31486.6 | 5.3 | 32.1 |
| Productivity | 22.1 | 5.2 | 0.1 | 2.5 |

Figures 4, 5, and 6 show histograms of the productivity variable in each dataset. Interestingly, we can observe that the productivity variable has a normal distribution with skewness very close to zero over all datasets. The productivity distribution of Dataset2 tends to be flatter than normal with a kurtosis value of 2.7. In contrast, the productivity of Dataset1 tends to be more sharply peaked than normal. A variable with a kurtosis value of less than three is less outlier-prone. The kurtosis values were less than three for all datasets, which ensures that the distribution of productivity is less outlier-prone. Although this is quite intuitive from the histograms, it was confirmed by the D'Agostino Pearson test for normality. We can also observe that the productivity of educational projects in Dataset2 is slightly smaller on average than industrial projects in Dataset1, and the distribution of productivity for Dataset2 is smaller than that of Dataset1. So it can be remarked that the projects in Dataset2 need less productivity, perhaps due to the type of projects that were developed in university as graduation projects, which are usually restricted by time and are less complex.

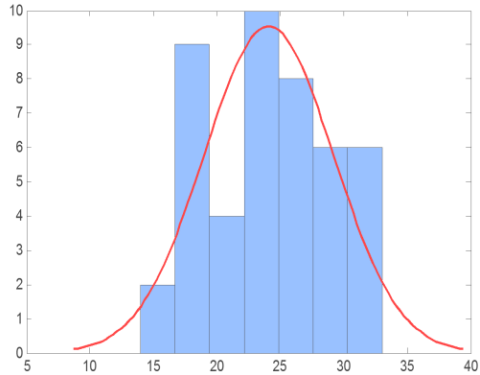

Figure 4. Productivity histogram of DS1

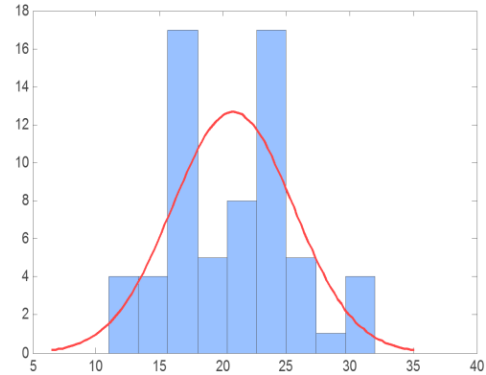

Figure 5. Productivity histogram for DS2

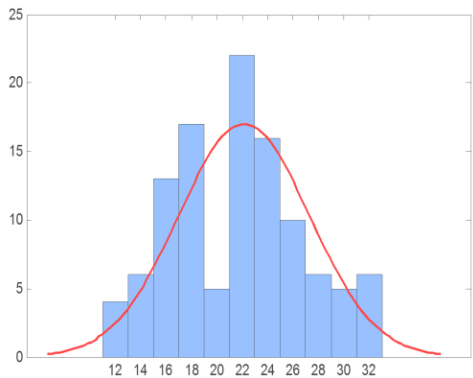

Figure 6. Productivity histogram of DS3

### 5.3 Evaluation measures

In this paper, we employ a set of reliable accuracy measures to evaluate the prediction model. These measures are unbiased and do not produce asymmetric error distribution as in Magnitude of Relative Error (MRE) and its derived measures: Mean Magnitude Relative Error and Performance indicator. The complete criticism about these biased measures can be found in [13][28]. However, the accuracy measures used in this paper are mainly derived based on Absolute Error (AE), see Equation 14. These measures are: Mean Absolute Error (MAE), Mean Balanced Relative error (MBRE), and the Mean Inverted Balanced Relative Error (MIBRE) as shown in Equations 15, 16, and 17, respectively.

In addition to that, we used other two recent accuracy measures proposed by Shepperd and MacDonell [28]; these are: Standardized Accuracy (SA) as shown in Equation 18, and effect size ($\Delta$) as shown in Equation 19. The SA metric is used mainly to check that the estimation model produces meaningful predictions that are better than a baseline of random guessing. The $\Delta$ is used to check that the predictions are not produced by chance. The predictor model is the one with a large SA and $\Delta$ [27]. Furthermore, we used the Wilcoxon sum rank test to check the significance difference between absolute errors of various estimation models. This statistical test has been used because the absolute errors are not normally distributed, so equivalent parametric tests such as t-test cannot be used.

$$AE_i = |y_i - \hat{y}_i| \tag{14}$$

$$MAE = \frac{\sum_i^n AE_i}{n} \tag{15}$$

$$MBRE = \frac{1}{n}\sum_1^n \frac{AE_i}{min(y_i, \hat{y}_i)} \tag{16}$$

$$MIBRE = \frac{1}{n}\sum_1^n \frac{AE_i}{max(y_i, \hat{y}_i)} \tag{17}$$

$$SA = 1 - \frac{MAE}{\overline{MAE}_{po}} \tag{18}$$

$$\Delta = \frac{MAE - \overline{MAE}_{po}}{SP_o} \tag{19}$$

Where:
$y_i$ and $\hat{y}_i$ are the actual and estimated effort of a project.
$SP_o$ is random guessing standard deviation.

$\overline{MAE}_{po}$ is "the mean value of a large number runs of random guessing, which means predicting a $\hat{y}_i$ for the target case $t$ by randomly sampling (with equal probability) over all the remaining $n$ - 1 cases and take $y_t = y$ where $r$ is drawn randomly from $1...n \wedge r \neq t$."

## 6. Results and Discussion

This section presents the empirical results obtained when we validated our proposed model and the compared models. We first validated all models using SA and effect size (Δ). The objective of this validation is twofold: 1) ensure that the models produce meaningful prediction better than random guessing, and 2) to ensure that all predictions are not generated by chance. Table 6 presents the results obtained over three datasets. A model with larger SA value indicates a reliable and meaningful prediction model. Also, a model with larger effect size indicates that the predictions are unlikely to have been generated by chance. From the table we can comfortably observe that the proposed model obtained larger SA over all datasets, whereas the previous models have larger SA over only Dataset1 and Dataset3. The quality of data in the original datasets (i.e., Dataset1 and Dataset2) contributed significantly to the obtained performance. The quality of data collected is affected by the experience of the estimator and the completeness of use case descriptions. For example, we can see that the experiences of students that is not mature enough and subject to a large degree of uncertainty have a great impact on the obtained performance over Dataset2. In addition, the time restriction on university projects force student to focus on the implementation phase more than the analysis phase, so we can find incomplete and vague use case diagram in most cases. However, the obtained SA results confirm the ability of our model to produce more meaningful prediction than random guessing, and to learn efficiently from a complex data structure. Nevertheless, the SA alone cannot give us the full picture about superiority of accuracy, therefore, we needed to consult other accuracy measures in order to build a final decision. The effect size test can tell us more about the meaningfulness of all prediction models. From Table 7 we can notice that effect sizes are considerably reasonable over Dataset1 and Dataset2, which suggest a good improvement against random guessing (i.e. Δ ≈ 0.5). Surprisingly, the effect size results on Dataset3 for all models were not high, perhaps due to the effect of outliers that is caused by merging both original datasets. From the initial observation, we can recommend that there should mainly be reliance on the datasets that come from organizations since their projects were usually collected by experts who have sufficient knowledge in objectory.

TABLE 6. SA and Δ analysis of all models, considering random guessing model as baseline

| Dataset | | Karner | S&W | Nassif | Our proposed Model |
|---|---|---|---|---|---|
| Dataset1 | SA% | 81.13 | 87.99 | 86.37 | 96.0 |
| | Δ | 0.44 | 0.48 | 0.47 | 0.53 |
| Dataset2 | SA% | 31.07 | 6.48 | 31.78 | 67.0 |
| | Δ | 0.44 | 0.09 | 0.45 | 0.58 |
| Dataset3 | SA% | 81.6 | 87.4 | 86.4 | 90.3 |
| | Δ | 0.28 | 0.3 | 0.3 | 0.49 |

Table 7 shows the amount of accuracy improvements in terms of SA and effect size on the proposed model, considering each previous model as the baseline. Interestingly, the proposed model generates good improvement over all datasets with good SA and reasonable effect size. For example, it produced great improvements against the S&W and Nassif model over Dataset1 and Dataset3, but fair improvements over Dataset2.

Table 7. SA and Δ analysis of our model, considering each previous models act as baseline

| Dataset | | Karner as baseline | S&W as baseline | Nassif as baseline |
|---|---|---|---|---|
| Dataset1 | SA% | 79.3 | 74.6 | 68.3 |
| | Δ | 0.51 | 0.43 | 0.44 |
| Dataset2 | SA% | 47.6 | 40.1 | 36.8 |
| | Δ | 0.49 | 0.35 | 0.34 |
| Dataset3 | SA% | 59.2 | 56.6 | 52.7 |
| | Δ | 0.44 | 0.43 | 0.43 |

Tables 8-10 show the obtained accuracy results with respect to MAE, MBRE, and MIBRE. These measures have been used because they behave very differently from each other, and they can effectively evaluate how well a model performs. The best model is the one with the minimum MAE, MBRE, and MIBRE. Remarkably, the results show that the proposed prediction model performed better than previous models over all datasets, which suggest that there are significant improvements over these other models.

TABLE 8. MAE results

| Dataset | Karner | S&W | Nassif | Our proposed Model |
|---|---|---|---|---|
| Dataset1 | 6120.62 | 3893.73 | 4419.96 | 1219.8 |
| Dataset2 | 329.04 | 446.42 | 325.64 | 201.1 |
| Dataset3 | 2698.3 | 1856.7 | 2000.6 | 1564.8 |

TABLE 9. MBRE results

| Dataset | Karner | S&W | Nassif | Our proposed Model |
|---|---|---|---|---|
| Dataset1 | 28.22 | 21.46 | 25.00 | 15.6 |
| Dataset2 | 23.18 | 30.36 | 24.28 | 17.4 |
| Dataset3 | 25.2 | 26.7 | 24.6 | 19.7 |

TABLE 10. MIBRE results

| Dataset | Karner | S&W | Nassif | Our proposed Model |
|---|---|---|---|---|
| Dataset1 | 20.72 | 16.67 | 18.47 | 11.8 |
| Dataset2 | 17.40 | 19.70 | 17.78 | 12.7 |
| Dataset3 | 18.8 | 18.5 | 18.1 | 15.1 |

To statistically justify the results obtained, we used the Wilcoxon sum rank statistical significance test based on absolute residuals, at the significance level of 0.05. The results of the statistical test are shown in Table 11. The results are denoted using square, triangle, and circle shapes to represent the significance among the models that were compared over a particular dataset. The presence of the square means that both compared models are statistically different over Dataset1, the presence of the triangle means that both compared models are statistically significant over Dataset2, and, finally, the presence of the circle means that both of the compared models are statistically significant over Dataset3. We can generally notice that proposed model generates statistically different and better predictions than all previous UCP models used in this paper.

TABLE 11. Statistical significance of test results over all datasets

|  | Karner |  |  |
|---|---|---|---|
| S&W | ■▲● | S&W |  |
| Nassif | ■  ● | ● | Nassif |
| Our proposed model | ■▲● | ■▲● | ■▲● |

The prediction model can be regarded as successful if it satisfies three criteria: accuracy, stability, and transparency. That means that the proposed model must be accurate in comparison with previous models in the same field. In this paper, we have seen that the proposed models outperformed all previous models using five accuracy measures. The construction model must be transparent and any researcher can trust and understand how the results were obtained. The construction of the proposed model is illustrated thoroughly in Section 4. Finally, the performance of the proposed model should be consistent when different evaluation measures are used. The performance of the proposed model was stable when different evaluation measures were used. To confirm these findings we used the Scott-Knott test analysis [8] The Scott-Knott is a multiple comparison statistical method based on the idea of clustering. The criteria of grouping is the significance test between the absolute errors of the method, using one-way analysis of variance. The Scott-Knott test requires the absolute errors to be transferred into normal distribution, so we used the Box-cox method to do so [8]. The graphical results of Scott-Knott test are shown in Figures 7, 8, and 9. The y-axis represents the transformed absolute errors and the small circles on each vertical line represent the mean of transformed absolute errors. The models are grouped together based on the significance tests; the model on the right hand side is the superior one. From the figures we notice that the proposed model is always located on the right hand side with a minimum of transformed absolute errors. Remarkably, the proposed model is not grouped with any other model, which confirms the performance of our model. However, we can see that S&W and Nassif behave similarly over Dataset1 and Dataset2 as they are grouped together in the same cluster.

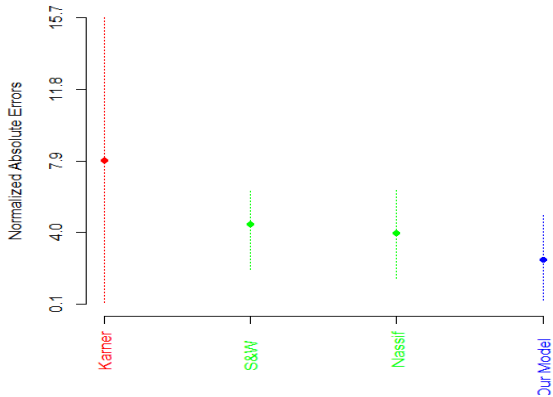

Figure 7. Plot of multiple comparison over Dataset1

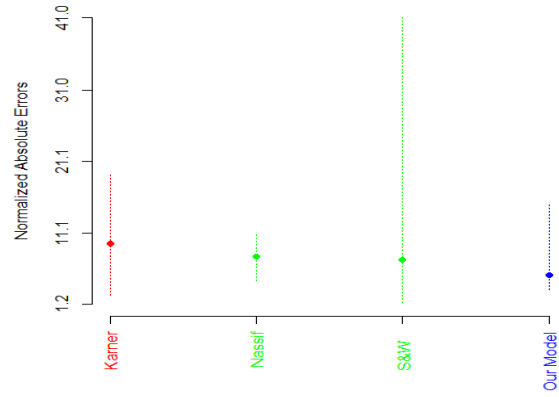

Figure 8. Plot of multiple comparison over Dataset2

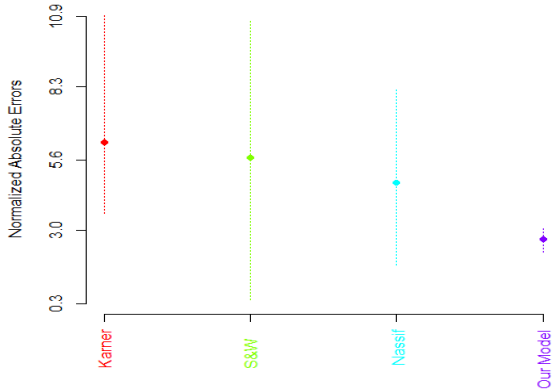

Figure 9. Plot of multiple comparison over Dataset3

## 7. Threats to Validity

This section describes the main threats to the validity of the study. First, we used the Bisecting k-medoids clustering algorithm for decomposing data into coherent groups. Bisecting the k-medoids enables us to avoid guessing the number of optimal classes and assigning a center of class from the instances in that class. We believe that the measure used is enough to give us an indication of how many clusters each training dataset needs. On the other hand, the number of neurons in RBFNN were selected empirically after several runs on each data separately. There is no established guideline on how to select the number of hidden neurons, so we used a trial and error approach to reach a compromise. Regarding evaluation measures, we used a set of reliable and unbiased measures (MAE, MBRE, and MIBRE) because the other common measures that are derived from MRE have been criticized as biased measures.

## 8. Conclusions

This paper puts forward a new model to support early effort estimation based on UCP size measure. The model attempted to fix some limitations in the previous studies by using a large number of observations and avoiding guessing productivity by predicting it from historical environmental factors. The large number of collected projects based on UCP will facilitate performing further investigations on the relationship between environmental factors and productivity to produce more accurate effort estimates. In our study, we built a hybrid model to predict and learn productivity and then to predict the effort from both UCP and productivity. To accomplish that, we used clustering and classification for the problem of productivity prediction, then we used RBFNN for predicting effort. The results obtained are encouraging and show significant improvements over previous models. The quality of data collection has a great effect on the performance of the models; we saw that all models work very well over Dataset1 and Dataset3 because they have data collected from software organizations.

Indeed, the enhancement achieved by the proposed model can be attributed to a number of reasons. Actually, most UCP datasets have a high granularity of labeling the data. Therefore the number of class label should take into consideration the structure of the training dataset. For example, the S&W and Nassif modes used only three and four class labels to classify productivity, which we believe is not sufficient. Even with suitable class labeling, inherent subclasses can be often discovered at a later stage. This is true in the software domain. For example, a practitioner can start with some class labels and then apply the decomposition technique on each class to get finer grained classes. The precision of class label can also be enhanced. However, the class labels obtained by the bisecting k-medoids algorithm was a fine-grained, labeled dataset. On the other hand, the use of RBFNN for effort prediction shows great improvement since the use of similarity based functions is a good choice of software data that often follow the same trend.

To conclude, we recommend excluding environmental factors from the UCP estimation process and making them available for the productivity prediction. Also, the use of non-linear prediction models, such as neural networks, tends to be more efficient than linear models.

## 9. Acknowledgement

Ali Bou Nassif would like to thank the University of Sharjah for supporting this work.